# HNAS-REG: HIERARCHICAL NEURAL ARCHITECTURE SEARCH FOR DEFORMABLE MEDICAL IMAGE REGISTRATION

*Jiong Wu and Yong Fan*

Center for Biomedical Image Computing and Analytics, Department of Radiology,
Perelman School of Medicine, University of Pennsylvania, Philadelphia, PA

## ABSTRACT

Convolutional neural networks (CNNs) have been widely used to build deep learning models for medical image registration, but manually designed network architectures are not necessarily optimal. This paper presents a hierarchical NAS framework (HNAS-Reg), consisting of both convolutional operation search and network topology search, to identify the optimal network architecture for deformable medical image registration. To mitigate the computational overhead and memory constraints, a partial channel strategy is utilized without losing optimization quality. Experiments on three datasets, consisting of 636 T1-weighted magnetic resonance images (MRIs), have demonstrated that the proposal method can build a deep learning model with improved image registration accuracy and reduced model size, compared with state-of-the-art image registration approaches, including one representative traditional approach and two unsupervised learning-based approaches.

## 1. INTRODUCTION

Deformable medical image registration plays a critical role in various medical image analysis applications, including medical image fusion [1], atlas building [2], and anatomical change quantification [3]. Recent advances in medical image registration have demonstrated that deep learning (DL) models built upon 3D convolutional neural networks (CNNs) can achieve promising image registration accuracy with reduced computation time [1, 4]. Most of the existing DL models for image registration are developed with manually designed network architectures, which are not necessarily optimal.

Due to the efficacy of U-Net in many medical image analysis tasks [5, 6], it becomes the first choice of network architecture in most deformable medical image registration studies [7]. To further improve its feature learning ability, some studies finetune the U-Net architecture by introducing vision transformer blocks [8]. Similar to the multi-resolution strategy of traditional registration methods, multi-resolution U-Nets have been developed to conduct the image registration in a coarse-to-fine manner [9].

On the other hand, neural architecture search (NAS) has been introduced into medical image segmentation studies, following its successful applications in general computer vision tasks, to facilitate optimal network architecture design by systemically searching for the best architectures automatically. For instance, a NAS-Unet method has been developed to search CNN operations in the down-sampling cell (DownSC) and up-sampling cell (UpSC) for medical image segmentation [10]. Although this method is able to identify a smaller network to achieve comparable segmentation results as a standard U-Net, its performance is limited since the method just searches the cell-level structures. To mitigate this problem, several studies enlarge the searching space by introducing network topology search in the NAS process. For instance, a multi-scale NAS has been developed to search both network backbone and cells for abdominal organ segmentation [11], and a differentiable neural network topology search method has been developed to find the optimal network topology to improve image segmentation performance [12].

Although NAS has achieved good performance in medical image segmentation studies, it has not been well studied for medical image registration. Recently, an automated learning approach has been developed to cooperate a searching of convolutional operations with a hyperparameter learning of the objective function for deformable medical image registration [13]. Similar to the NAS method for image segmentation [10], its performance is also constrained by the limited searching space since only convolutional operations are searched. In this study, we develop a hierarchical NAS method, consisting of network topology search and convolutional operations search, to identify the optimal network architecture for medical image registration. To mitigate the computational overhead and memory constraints, we introduce a partial channel strategy in the proposed framework. To our best knowledge, this is the first piece of work that achieves automated architecture searching for medical image registration by searching the architecture in both the network cell and topology levels.

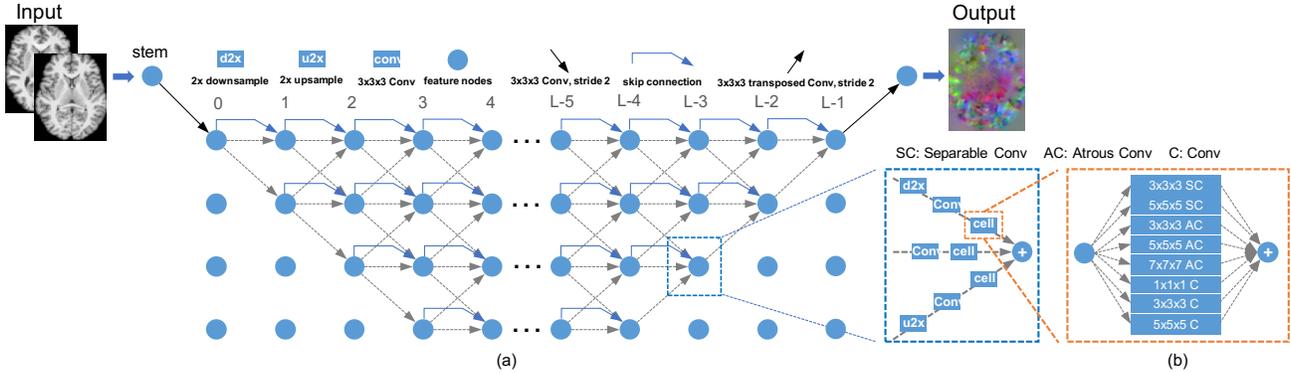

**Fig. 1**. Overview of the architecture search space for deformable medical image registration: (a) Topology search space for HNAS-Reg; (b) Cell search space for HNAS-Reg.

## 2. METHODS

### 2.1. Problem Formulation

Given a pair of source image $I_0$ and target image $I_1$, DL-based deformable image registration methods aim to use 3D CNNs to predict a spatial deformation field, $\phi$, which can be parameterized with a stationary velocity field $v$, via $\phi = \int_0^1 v_\tau\, dt$, such that $I_0 \circ \phi^{-1}$ is well-aligned to $I_1$. Denoting all unknown network parameters by $\boldsymbol{\theta}$, the 3D CNNs can be optimized to approximate $v = f_{\boldsymbol{\theta}}(I_0, I_1)$, and the DL model's objetivate function for image registration is formulated as

$$\mathcal{L} = \min_{v} \mathcal{L}_{sim}(I_0 \circ \phi^{-1}, I_1) + \lambda \mathcal{L}_{smooth}(v), \quad (1)$$

where $\mathcal{L}_{sim}(\cdot,\cdot)$ denotes an image similarity term for quantifying the difference between $I_0 \circ \phi^{-1}$ and $I_1$, $\mathcal{L}_{smooth}(\cdot)$ denotes a regularization term used to ensure the smoothness of $v$, and $\lambda$ is a weighting factor to balance the image similarity and the regularization term.

To facilitate the network architecture search we introduce in a DL network a parameter $\boldsymbol{\alpha}$ representing the weights of a series of convolutional operations in the cells and a network topology parameter $\boldsymbol{\eta}$ representing the weights of edges between linked feature nodes. The objective function of our proposed hierarchical search framework is formulated as

$$\begin{aligned}
&\min_{\boldsymbol{\alpha},\boldsymbol{\eta}} \mathcal{L}_{val}(\boldsymbol{\alpha}, \boldsymbol{\eta}, \boldsymbol{\theta}^*; I_0, I_1), \\
s.t. \quad &\boldsymbol{\theta}^*(\boldsymbol{\alpha},\boldsymbol{\eta}) = \underset{\boldsymbol{\theta}}{argmin}\ \mathcal{L}_{tr}(\boldsymbol{\theta}; \boldsymbol{\alpha}, \boldsymbol{\eta}, I_0, I_1).
\end{aligned} \quad (2)$$

Both $\mathcal{L}_{val}$ and $\mathcal{L}_{tr}$ have a similar form as the Eq. (1). The first level of network architecture search optimizes the network parameters $\boldsymbol{\theta}$ by fixing the network architecture parameters $\boldsymbol{\alpha}$ and $\boldsymbol{\eta}$ on a training dataset, and the second level of optimization trains the network architecture parameters $\boldsymbol{\alpha}$ and $\boldsymbol{\eta}$ by fixing $\boldsymbol{\theta}$ using a validation dataset. Such two-level search processes are optimized jointly to search for the optimal network architecture for image registration.

### 2.2. Hierarchical Architecture Search Space

Inspired by successful NAS techniques [12, 14], we propose a network topology search space with fully connected edges between adjacent layers and resolutions as illustrated in Fig. 1 (a). The search space has $L$ layers and each layer consists of $N = 4$ feature nodes (blue nodes) at different spatial resolutions. Three types of blocks, including up-sampling $UpS(\cdot)$, no-sampling $SameS(\cdot)$, and down-sampling $DownS(\cdot)$, are assigned on upsample, nosample, and downsample edges, respectively. Each block consists of a feature map scaling layer, a regular convolutional layer, and cells of convolutional operations. To facilitate an efficient search process and reduce the search space, inspired by the successful network architectures for medical image registration, the architecture is designed to have a similar U-shape encoder-decoder. In addition, skip connections are built between adjacent cells in the same spatial resolution level to improve the feature learning capability. In total, $E = 10L - 40$ edges exist in different paths from the input image pair to the output velocity field in the topology search space.

We utilize continuous relaxation algorithms to convert the discrete network topology search problem into a continuous one. Let $x_s^l$ denotes the output feature maps of a node in the $l$-th layer with the feature map resolution of $s$, we parameterize the weights of three edges connected to this node as $\eta_{\frac{s}{2}\to s}^{l-1}, \eta_{s\to s}^{l-1}$ and $\eta_{2s\to s}^{l-1}$, then the $x_s^l$ can be calculated as

$$\begin{aligned}
x_s^l &= \eta_{\frac{s}{2}\to s}^{l-1} UpS\left(x_{\frac{s}{2}}^{l-1}; \boldsymbol{\alpha}\right) + \eta_{s\to s}^{l-1} SameS(x_s^{l-1}; \boldsymbol{\alpha}) \\
&+ \eta_{2s\to s}^{l-1} DownS(x_{2s}^{l-1}; \boldsymbol{\alpha}).
\end{aligned} \quad (3)$$

The scalars $\eta$ are normalized through SoftMax such that

$$\eta_{\frac{s}{2}\to s}^{l-1} + \eta_{s\to s}^{l-1} + \eta_{2s\to s}^{l-1} = 1, \quad \forall s, l$$

$$\eta_{\frac{s}{2}\to s}^{l-1} \geq 0, \quad \eta_{s\to s}^{l-1} \geq 0, \text{ and } \eta_{2s\to s}^{l-1} \geq 0, \quad \forall s, l. \quad (4)$$

For the cell-level search space, as shown in Fig. 1 (b), each type of cells has a set of basic convolutional operations where the numbers and the resolutions of the input and output feature maps are the same. Our algorithm searches the operations in each type of cells independently, with one operation selected from the following:

- $3 \times 3 \times 3$ Separable Conv
- $5 \times 5 \times 5$ Separable Conv
- $3 \times 3 \times 3$ Atrous Conv
- $5 \times 5 \times 5$ Atrous Conv
- $7 \times 7 \times 7$ Atrous Conv
- $1 \times 1 \times 1$ Conv
- $3 \times 3 \times 3$ Conv
- $5 \times 5 \times 5$ Conv

To facilitate the cell-level search procedure, we follow the idea of continuous relaxation in NAS-Unet [10] in our method. Moreover, to mitigate the computational overhead and memory constraints, we embedded a partial channel strategy in the proposed framework. Specifically, the input feature maps of each cell $x_{in}$ are separated into $k$ groups, and only one group is randomly selected and calculated by the cells. Let $o(\cdot)$ denotes one convolutional operation of the candidate set $\mathcal{O}$, a column vector $\alpha_o^{i,j}$ (with the size of $|\mathcal{O}|$) denotes the weights of candidate operations, the output of each cell can be calculated as:

$$\hat{o}(x_{in}; K_{ij}) = \sum_{o \in \mathcal{O}} \frac{exp(\alpha_o^{ij})}{\sum_{o' \in \mathcal{O}} exp(\alpha_{o'}^{ij})} \cdot o(K_{i,j} * x_{in}) \oplus (1 - K_{i,j}) * x_{in}, \quad (5)$$

where $K_{i,j}$ denotes channel sampling mask and $\oplus$ denotes concatenation operation.

### 2.3 Optimization and Decoding

With the definition of search spaces and parameter relaxation, the architecture search is converted to a continuous optimization problem. However, such conversion may result in a "discretization gap" in the search space, leading to a suboptimal searched architecture [12]. To alleviate this problem, we introduce the entropy losses to the binarization of $\boldsymbol{\eta}$ and $\boldsymbol{\alpha}$:

$$\mathcal{L}_\eta = \langle \boldsymbol{\eta}, log(\boldsymbol{\eta}) \rangle, \quad \mathcal{L}_\alpha = \langle \boldsymbol{\alpha}, log(\boldsymbol{\alpha}) \rangle. \quad (6)$$

Since the spatial deformation field is parameterized with a stationary velocity field in our study, our method is capable of obtaining both the forward and backward deformation fields. Therefore, we formulate our optimization based on both the deformation field $\phi_{I_0 I_1}$ and inverse deformation field $\phi_{I_1 I_0}$. Then the objective function of the proposed hierarchical NAS method for image registration can be reformulated as

$$\min_{\alpha, \eta} \mathcal{L}_{val}(\boldsymbol{\alpha}, \boldsymbol{\eta}, \boldsymbol{\theta}^*; I_0, I_1) + \gamma \mathcal{L}_\alpha + \beta \mathcal{L}_\eta$$
$$s.t. \quad \boldsymbol{\theta}^*(\boldsymbol{\alpha}, \boldsymbol{\eta}) = \underset{\boldsymbol{\theta}}{argmin} \mathcal{L}_{tr}(\boldsymbol{\theta}; \boldsymbol{\alpha}, \boldsymbol{\eta}, I_0, I_1), \quad (7)$$

where $\mathcal{L}_{val/tr} = \min_v \mathcal{L}_{sim}(I_0 \circ \phi_{I_0 I_1}, I_1) + \mathcal{L}_{sim}(I_1 \circ \phi_{I_1 I_0}, I_0) + \lambda \mathcal{L}_{smooth}(v)$, and $\gamma$ and $\beta$ denote the weights to balance the loss terms of $\mathcal{L}_{val}, \mathcal{L}_\alpha$ and $\mathcal{L}_\eta$. After the architecture search, only one path from the input image pair to the output velocity field is selected by using the *Dijkstra* algorithm. For each type of cell, only one convolutional operation with the maximum probability is selected.

## 3. EXPERIMENTS

### 3.1 Datasets and Experiment Setup

We evaluated our method on two public brain imaging datasets including (a) MICCAI 2012 Multi-Atlas Labelling Challenge (MALC) dataset[1] consisting of 35 T1-weighted images with 134 manually delineated brain structures, and (b) Mindboggle-101 dataset[2] consisting of 101 T1-weighted images with 50 manually delineated cortical structures. These images were used for testing only. We randomly selected 500 T1 brain MR images from ADNI 1 cohort[3] and divided them into 400 and 100 images for training and validation. Preprocessing steps, including skull-stripping, intensity normalization, and spatial alignment using 12 parameters of affine registration, were conducted using FreeSurfer [15]. All the images were resampled at a resolution of $1 \times 1 \times 1~mm^3$ and center cropped with a size of $144 \times 160 \times 192$.

A total of 150,000 epochs with a batch size of 1 was adopted in both the architecture search and training procedure, and Adam optimizer was employed. We set the initial learning rates of 0.001 and 0.0001 for architecture search and training, respectively. The hyper-parameters $L, \lambda,$ and $\beta$ were respectively set to be 8, 0.5, 0.2, and 0.1, and normalized cross-correlation (NCC) was used as the metric in similarity term. We used the Dice score on 34 and 100 pairs of registration results from MALC and Mindboggle-101 to evaluate the registration performance. A Wilcoxon signed-rank test was performed to quantify the significance of the difference between different methods.

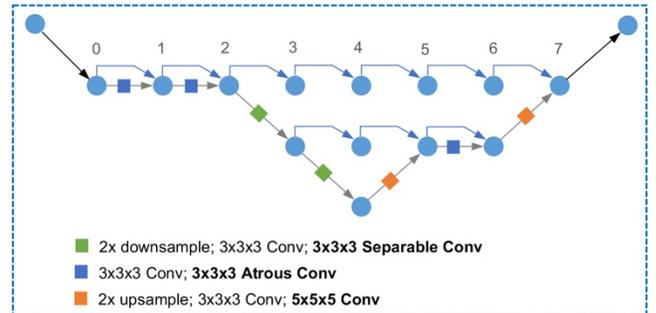

**Fig. 2**. The final architecture of HNAS-Reg, bold font indicates the searched convolutional operations in each type of cells.

---

[1] http://www.neuromorphometrics.com/2012_MICCAI_Challenge_Data.html
[2] https://mindboggle.info/data#
[3] http://adni.loni.usc.edu/

We compared the DL model built by the proposed method, referred to as HNAS-Reg, with four state-of-the-art representative medical image registration methods: 1) one conventional registration method SyN [16]; 2) two unsupervised DL-based registration methods including a diffeomorphic variant of Voxelmorph (VM-diff) [7] and fast symmetric diffeomorphic image registration (FSD-Reg) [17]; and 3) a U-Net-based registration method (UNet-Reg) in which the U-Net was used to predict the velocity field and the other settings were the same as the searched architecture of HNAS-Reg. All the DL models were trained and evaluated on the same training, validation, and testing datasets.

### 3.2 Experimental Results

Fig. 2 shows the optimal path and optimal convolutional operations of the three types of cells identified by the proposed method. Mean and standard deviations of the Dice scores across all manually delineated structures between images registered by the five methods under comparison on MALC and Mindboggle-101 summarized listed in Table 1. The Dice scores of HNAS-Reg were significantly higher than those obtained by the other four methods, with a $p$-value of 0.05, and HNAS-Reg had the smallest model size of 0.820 MB.

**Table 1**. The mean and standard deviations of the Dice scores across all manually delineated structures on MALC and Mindboggle-101, as well as model size on SyN, VM-diff, FSD-Reg and HNAS-Reg. Bold font indicates statistically significant group difference.

|  | MALC | Mindboggle-101 | # (MB) |
|---|---|---|---|
| SyN | 0.591 (0.030) | 0.534 (0.019) | - |
| VM-diff | 0.588 (0.030) | 0.543 (0.023) | 1.016 |
| FSD-Reg | 0.589 (0.030) | 0.555 (0.020) | 1.124 |
| UNet-Reg | 0.592 (0.029) | 0.558 (0.020) | 1.118 |
| HNAS-Reg | **0.602 (0.029)** | **0.575 (0.020)** | 0.820 |

To prove the effectiveness of the network topology search component, we set network topology to be the U-Net and only conduct convolutional operations search based on the blocks of $DownS(\cdot)$ and $UpS(\cdot)$ of HNAS-Reg. After optimization, we obtained an optimal architecture in which $3 \times 3 \times 3$ *Conv* and $5 \times 5 \times 5$ *Conv* respectively selected in $DownS(\cdot)$ and $UpS(\cdot)$ with the model size of 1.100 MB. Dice scores of this model on MALC and Mindboggle-101 were respectively 0.573 and 0.538, worse than those obtained by the proposed method.

### 4. CONCLUSION

We have proposed a hierarchical network architecture search framework to search both network topology and convolutional operations for deformable image registration. The computational overhead and memory constraints associated with the NAS are mitigated considerably by a partial channel strategy. Comparison results have demonstrated the superiority of the proposed method and the effectiveness of the network topology search component.

### 5. COMPLIANCE WITH ETHICAL STANDARDS

This study was conducted using publicly available data.